\documentclass[sigconf]{acmart}
\usepackage{tikz}
\usepackage{amsmath}
\usepackage{microtype}
\usepackage{graphicx}
\usepackage{multicol}
\usepackage{xspace}
\usepackage{xcolor}
\usepackage{booktabs} %
\usepackage{adjustbox}
\usepackage{hyperref}
\usepackage{colortbl}
\usepackage{subcaption}
\usepackage{enumitem}
\usepackage{tcolorbox}

\AtBeginDocument{%
  }

\acmYear{2024}\copyrightyear{2024}
\setcopyright{rightsretained}
\acmConference[FlexScience '24]{Workshop on AI and Scientific Computing at Scale using Flexible Computing Infrastructures}{June 3--4, 2024}{Pisa, Italy}
\acmBooktitle{Workshop on AI and Scientific Computing at Scale using Flexible Computing Infrastructures (FlexScience '24), June 3--4, 2024, Pisa, Italy}
\acmDOI{10.1145/3659995.3660038}
\acmISBN{979-8-4007-0642-4/24/06}

\newtcolorbox{observationbox}{colback=red!5!white,colframe=red!75!black}

\begin{document}
\title{Breaking the Memory Wall: A Study of I/O Patterns and GPU Memory Utilization for Hybrid CPU-GPU Offloaded Optimizers}

\author{Avinash Maurya}
\affiliation{
    \institution{Rochester Institute of Technology}
    \city{Rochester, NY}
    \country{USA}
}
\email{am6429@cs.rit.edu}

\author{Jie Ye}
\affiliation{
    \institution{Illinois Institute of Technology}
    \city{Chicago, IL}
    \country{USA}
}
\email{jye20@hawk.iit.edu}

\author{M. Mustafa Rafique}
\affiliation{
    \institution{Rochester Institute of Technology}
    \city{Rochester, NY}
    \country{USA}
}
\email{mrafique@cs.rit.edu}

\author{Franck Cappello}
\affiliation{
    \institution{Argonne National Laboratory}
    \city{Lemont, IL}
    \country{USA}
}
\email{cappello@anl.gov}

\author{Bogdan Nicolae}
\affiliation{
    \institution{Argonne National Laboratory}
    \city{Lemont, IL}
    \country{USA}
}
\email{bnicolae@anl.gov}
\renewcommand{\shortauthors}{Avinash Maurya et al.}

\begin{abstract}
Transformers and LLMs have seen rapid adoption in all domains. Their sizes have exploded to hundreds of billions of parameters and keep increasing. Under these circumstances, the training 
of transformers is slow and often takes in the order of weeks or months. Thanks to 3D model parallelism (data, pipeline, and tensor-level parallelism), the training can scale to a large number of GPUs, which reduces the duration of the training but dramatically increases the
cost. Even when a large number of GPUs are available, the aggregated GPU memory is often not enough to hold the full training state (optimizer state, model parameters, and gradients). To compensate, state-of-the-art approaches offload the optimizer state at least partially to the host memory and perform hybrid CPU-GPU computations. Such flexible solutions dramatically reduce the 
GPU memory utilization, which makes it feasible to run the training on a smaller number
of GPUs at the cost of performance penalty. Unfortunately, the challenges and bottlenecks of
adopting this strategy are not sufficiently studied by state-of-the-art, which results in 
poor management of the combined host-GPU memory and poor overlapping between data movements and computations. In this paper, we aim to fill this gap by characterizing the behavior of
offloaded training using the DeepSpeed runtime. Specifically, we study the GPU
memory utilization over time during each iteration, the activity on the PCIe related
to transfers between the host memory and the GPU memory, and the relationship between
resource utilization and the steps involved in each iteration. Thanks to this study, we
reveal opportunities for future improvements of offloading solutions, which enable greater
flexibility to optimize the cost-performance trade-off in the context of transformer and
LLM training.
\end{abstract}




\maketitle

\section{Introduction}
\label{sec:intro}

Transformers and Large-Language Models~(LLMs) have seen increasing 
adoption in various domains ranging from scientific research to
industrial applications~\cite{zhao2023survey}. While traditionally used for creative 
text generation, prompt completion, comprehension, and summarization,
these learning models are successfully tackling multi-modal
data sources, thanks to cross-attention~\cite{MultiModal-Transformers23}.
Additionally, recent initiatives, such as LLMs for science (e.g., DeepSpeed4Science\cite{DeepSpeed4Science23}) are beginning to explore use cases that involve specialized domain-specific languages for tasks such as genome sequencing, protein structure prediction, equilibrium distribution prediction, etc. The versatility and democratization of LLMs have led to unprecedented scale of development and discovery across multiple fields.

In a quest to improve the quality, LLMs are routinely made of billions of parameters with models such as GPT-3, Llama-2-70b, and BLOOM requiring hundreds of gigabytes of GPU memory just to store the model parameters~\cite{workshopBLOOM176BParameterOpenAccess2023,touvronLlamaOpenFoundation2023}. Several predictions anticipate LLMs will reach trillion-scale 
parameters in the near future, e.g., Google Switch-C (1.6T)~\cite{google-switch}, WuDao 2.0 (1.75T)~\cite{zeng2022glm}, and M6-10T~\cite{lin2021m6}. 
Despite advances in technologies that enable LLM training to scale (hybdrid data-, pipeline- and tensor parallelism, sharding of model parameters and optimizer state, layout and communication optimizations, etc.), the rapid growth in the number of parameters has outpaced the available GPU memory, creating a significant `memory wall' that makes it challenging to train and run these massive models efficiently~\cite{MemWall23,PagedAttention}.

\begin{figure*}[t]
    \centering
    \includegraphics[width=\linewidth]{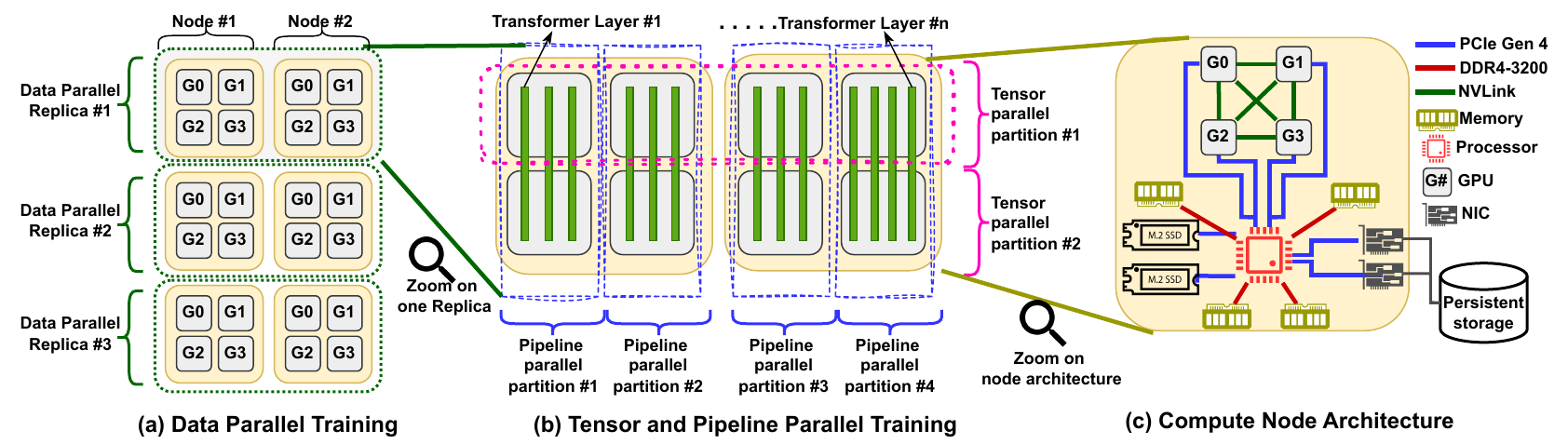}
    \caption{Data, pipeline, and tensor parallel runtime training. Compute node configuration consisting of 4 A100-40GB GPUs.}
    \label{fig:3d-parallel}
    \Description{
    A diagram that illustrates data, pipeline and tensor parallelism
    mapped on a typical compute node architecture consisting of four A100-40GB GPUs.}
\end{figure*}

Specifically, users face an important trade-off: either they pay a high cost,
by sharding the models at fine granularity on a large number of GPUs, such
that the aggregated GPU memory is sufficient to hold all necessary data
structures (model parameters, gradients, optimizer state, activations), or
they run on a smaller number of GPUs and accept a performance penalty,
by using offloading solutions, such as DeepSpeed 
ZeroInfinity~\cite{ZeroInfinity-SC21}, which move a part of the data 
structures and computations to the host memory and the CPU 
respectively
(notably the optimizer state and the updates of the model parameters
based on it). While the flexibility to exploit
offloading at least partially is well acknowledged~\cite{ren2021zero,deepspeed-offloadapp}, the emergent
challenges and opportunities in this space are not well understood.
With increasing focus on energy efficiency and cost-effectiveness, there
is an acute need for more insight in this direction.

In this paper, we aim to fill the aforementioned gap. We propose to characterize 
the behavior of offloading solutions at fine granularity, in order to derive
patterns and opportunities that can be exploited to design better offloading 
strategies. We specifically consider the case of DeepSpeed 
ZeroInfinity~\cite{ZeroInfinity-SC21}, but our methodology can be easily adapted
to other LLM and transformer training runtimes. We summarize our contributions below:

\begin{enumerate}%

\item We discuss how state-of-the-art LLM training runtimes break down each
training iteration into intermediate steps, and explain how these steps
interleave to enable efficient overlap between the computations and 
communications (\S~\ref{sec:background}).

\item We run extensive experiments that measure the utilization of GPU memory 
and PCIe links (used for data transfers between the host memory and the 
memories of the GPUs) over time during each training iteration 
and correlate the utilization with the steps of the training iteration.
We observe significant fluctuations in utilization that lead to poor combined 
host-GPU memory management and suboptimal overlapping between data movements 
and computations (\S~\ref{sec:characterize}).

\item We reveal an opportunity to recycle large amounts of GPU memory during
the backward pass (corresponding to activations that are not needed after
the gradient computations) in order to buffer and asynchronously flush 
a part of the gradients to the host memory. Furthermore, we reveal
an opportunity to accelerate the update of the model parameters after the
backward pass, by temporarily offloading the optimizer state partially back
to the GPUs (\S~\ref{sec:conclusions}).

\end{enumerate}

\section{Background and Related Work}
\label{sec:background}

\textbf{Data parallelism:} 
Data parallelism is the most widely used technique to accelerate the training 
of deep learning models~\cite{DBLP:journals/pvldb/LiZVSNLPSVDC20}. It creates replicas of 
the learning model on multiple workers, each of which is placed on a different device and/or
compute node, as illustrated in Figure~\ref{fig:3d-parallel}(a). The input data is randomly shuffled and partitioned among the workers at each epoch. 
During the forward pass, the workers simply process their
mini-batches from the partition of their dataset in an embarrassingly 
parallel fashion. Then, during the backward pass, the model parameters 
are updated based on the average gradients of all replicas (instead of the
local gradients), which effectively synchronizes all replicas to learn
the same patterns from all partitions. Data parallelism leads to accelerated training 
because the partitioning of the input data results in fewer iterations per epoch. 
Furthermore, the redundancy of replication and the pressure it creates on memory
utilization can be eliminated by partitioning the data structures across
the GPUs at the expense of higher communication overheads to access remote 
partitions during computations (all-gather). A prominent example
is ZeRO-1/2/3~\cite{rajbhandari2020zero}, which partitions the optimizer states,
gradients, and model parameters respectively on the GPUs.

\textbf{Pipeline and tensor parallelism:}
Pipeline and tensor parallelism are two complementary techniques that enable the training of large learning models 
that do not fit in the memory of a single GPU. Pipeline parallelism leverages the 
idea that learning models can be split into stages, each of which can placed on
a separate GPU. Then, the forward and backward pass corresponding to different 
mini-batches can be overlapped by activating all stages in parallel, at the cost
of caching more activations~\cite{korthikanti2023reducing} for the backward pass and the extra communication
overheads needed to transfer the outputs between the stages on different GPUs.
Tensor parallelism leverages the idea that even individual layers and tensors can
be sharded and distributed horizontally across multiple GPUs, at the expense of
incurring extra communication to synchronize the computations over the same
tensor. A prominent example is DeepSpeed Megatron~\cite{shoeybi2019megatron,megatron-deepspeed}. The combination of
pipeline and tensor parallelism and how it maps to a typical HPC compute node
is illustrated in Figures~\ref{fig:3d-parallel}(b) and~\ref{fig:3d-parallel}(c), respectively.

\textbf{Decoupled update phase to enable gradient accumulation:}
The model parameters are updated based on the optimizer state and the gradients at the end of each forward and backward pass in traditional deep learning model training.
However, the adoption of data, tensor, and pipeline parallelism introduces expensive synchronization points necessary for the updates, e.g., gradient all-reduce in the case of data parallelism. As a consequence, a frequent optimization is \emph{gradient accumulation}, where the gradients are summed up over multiple forward and backward passes, then they are averaged, and the average gradients are used in a separate update step. In our study, when gradient accumulation is used, we refer to the last forward and backward pass before the update of the model parameters as the \emph{gradient boundary} pass. By contrast, any forward or backward pass that is not followed by a model parameter update is referred to as \emph{non-gradient boundary}.

\begin{table}[t]
    \caption{Model and optimizer state sizes for different models.}
    \label{tab:model-size}
    \centering
    \begin{tabular}{|l||c|c|c|c|c|c|c|c|}
        \hline
         Model size (Billions) & 7B & 13B & 30B & 70B & 130B & 175B \\
         \hline \hline
         Model states (GB) & 24 & 46 & 120 & 240 & 473 & 640 \\
         Optimizer states (GB) & 96 & 187 & 475 & 960 & 1890 & 2572 \\
         \hline
    \end{tabular}
\end{table}

\textbf{Host-offloaded update phase using mixed precision:}
With the model and optimizer state sizes exploding (Table~\ref{tab:model-size})~\cite{workshopBLOOM176BParameterOpenAccess2023,touvronLlamaOpenFoundation2023,zeng2022glm},
approaches such as DeepSpeed ZeroInfinity~\cite{ZeroInfinity-SC21}, CoTrain~\cite{li2023cotrain}, etc. have explored 
the idea of moving large data structures required during training to the host memory, 
notably the optimizer state. This makes it feasible to train LLMs on much smaller 
aggregated GPU memory footprint, albeit at the cost of performance penalty.
Specifically, by keeping a master copy of the optimizer state and 
model parameters on the host memory in high 32-bit floating point (FP32) precision, the forward
pass and backward pass can operate with model parameters in lower 16-bit floating point (FP16)
precision to calculate FP16 gradients, which are then flushed to the
host memory and upscaled to FP32 precision. Then, the update of the
parameters can proceed directly on the CPU and a downscaled FP16 copy
can be transferred to the GPUs for the next iteration. In this case, an
important bottleneck is the I/O bandwidth between the host memory and the 
memory of the GPUs, which is limited by PCIe links. This bottleneck is further 
exacerbated by competition for the PCIe links due to inter-node communication 
needed to implement tensor, pipeline, and data parallelism, which results in additional 
overheads during the forward pass (waiting for the copy of model parameters from the 
host to the GPU) and the backward pass (waiting to flush the gradients from
the GPU to the host). Another important bottleneck is the
low computational capability of the CPUs, which are orders of magnitude
slower than the GPUs. Under such circumstances, despite being 
simple and embarrassingly parallel, the operations involved in updating 
the model parameters and the optimizer state lead to a significant
runtime overhead, which otherwise is negligible when running them on 
the GPUs.

\begin{figure}[t]
    \centering
    \includegraphics[width=0.8\columnwidth]{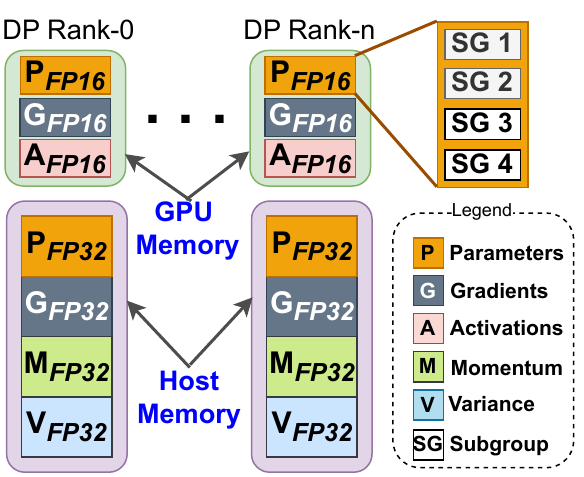}
    \caption{Subgroup sharding of model parameters, gradients, and activations on GPUs.}
    \label{fig:subgroup-partitioning}
    \Description{
    A diagram that illustrates subgroup sharding of model parameters, gradients, and activations across GPU and host memory with DeepSpeed ZeRO-3 settings.}
\end{figure}

\textbf{Sharding of partitions into subgroups:}
When the partitions corresponding to the model parameters, gradients, and activations
are so large that they do not fit on the GPU memory even at low FP16 precision, a
common mitigation strategy is sharding the partition into subgroups. Specifically,
the forward pass can serially process each subgroup one after another, while 
accumulating (and/or checkpointing) the activations. Then, during the backward pass,
the same process can be repeated to compute the gradients. If the subgroup is large
enough to saturate the GPU compute capability, the serial processing of multiple
subgroups incurs minimal overhead compared with processing the whole partition
at once, but with the added benefit of reduced memory utilization. This strategy
is illustrated in Figure~\ref{fig:subgroup-partitioning}.

\textbf{Hybrid GPU-CPU partitioning and update phase processing:}
Once the gradients have been computed by the backward pass, model parameters are updated that involve embarrassingly parallel computations
since each model parameter only depends on its own corresponding gradient and optimizer state, e.g., momentum and variance in the case of Adam~\cite{ZeroInfinity-SC21}. 
This property is exploited by techniques such as ZeRO-Offload++ \cite{deepspeed-offloadapp} in order to partition the optimizer state between the CPU and GPU, instead of completely offloading the optimizer state to the CPU, which 
enables the update step to proceed in parallel. However, such techniques 
require that the model parameters and optimizer state are pre-partitioned
on the CPU and GPU using a fixed ratio 
that is determined at the start of the training process.
The problem of how to choose the fixed ratio is difficult, because
there is a fine balance between avoiding out-of-memory (OOM) errors on
the GPUs, while at the same time maximizing the GPU memory utilization
to improve performance and scalability. With increasing mini-batch
size and therefore activation sizes, a fixed ratio results in
a majority of the optimizer state ending up on the CPU, thereby negating
the benefits of hybrid computations.

\begin{figure*}[t]
    \centering
    \includegraphics[width=0.99\linewidth]{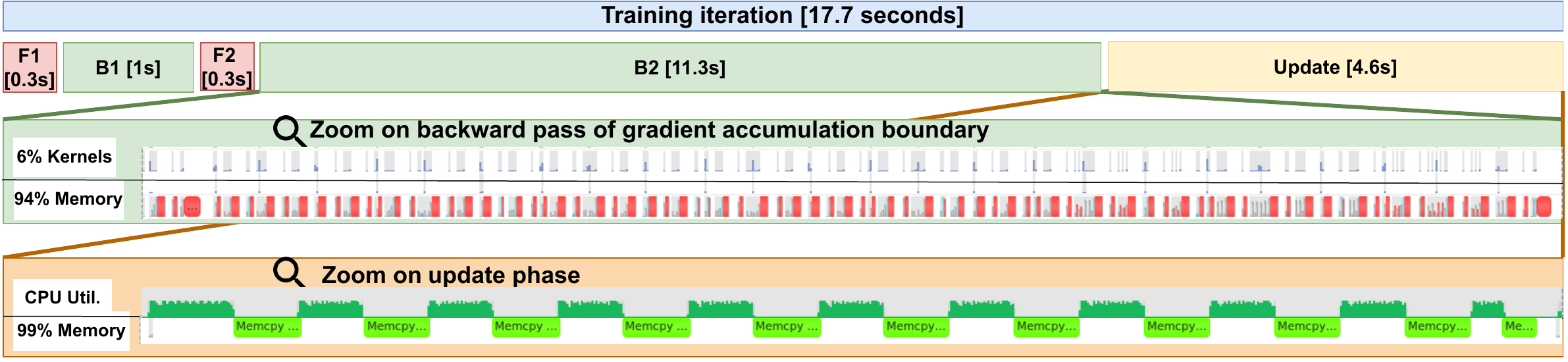}
    \caption{Breakdown of different training stages for 13B model running on 1 GPU with TP=1, DP=1, micro-batch=1, gradient accumulation=2, ZeRO stage=3. The optimizer is completely offloaded to the CPU.}
    \label{fig:iter-time-breakdown}
    \Description{
    A diagram that illustrates the breakdown time of GPU kernel, CPU utilization, and memory accesses across device and host due to data transfer.}
\end{figure*}

\textbf{Motivation of the study:}
The sheer complexity of modern training runtimes for LLMs and transformers,
as summarized above, makes it unfeasible to study all combinations of optimizations. However, 
some observations
can be made to
narrow down the scope of the study. First, the training process is iterative
in nature, which means the behavior is repetitive and therefore it
is sufficient to study a single iteration in detail. Second, each iteration
consists of one or more forward and backward passes, followed by an update
step. Thus it is important to study the differences in the behavior 
as the iterations transition between the steps. Intuitively, we expect fluctuations in resource utilization that prompt the need for flexible 
solutions to adapt the offloading to the training behavior.
Third, to enable efficient hybrid CPU-GPU partitioning and processing of the update 
phase, it is important to focus on several important aspects:
(1) the memory utilization of the GPUs as the iteration progresses in
time, which determines not only how much of the optimizer state and 
model parameters need to be offloaded to the host memory, 
but also \emph{when}; (2) the utilization of the PCIe links that
are stressed by the constant data movement between the host memory
and the GPU memory; and (3) the sizes of the tensors involved in the
data movements, which enables us to reason about how to overlap
data movements with computations and how to mitigate competition
for host memory and PCIe links, e.g., by initiating the transfer
of large tensors restricts the granularity of the future scheduling 
decisions due to keeping resources busy for longer than the transfer
of smaller tensors. Based on these observations, we narrowed the
scope of our study accordingly.

\section{Study of I/O Patterns and GPU Memory Usage}
\label{sec:characterize}

\begin{figure*}[t]
\begin{subfigure}[t]{0.32\linewidth}
    \centering
    \includegraphics[width=\linewidth]{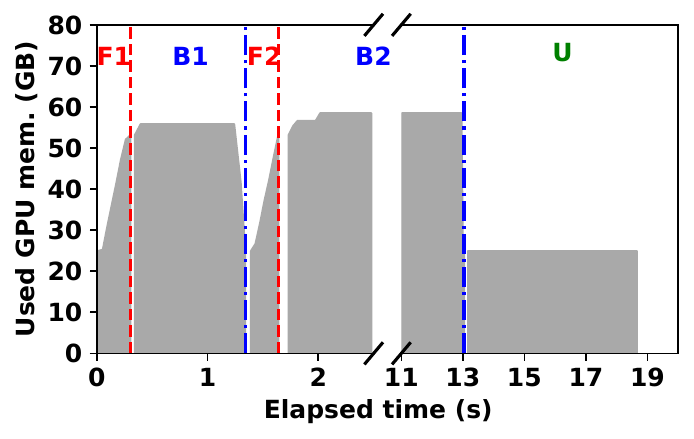}
    \caption{GPU memory utilization.}
    \label{fig:gpu-mem-util}
\end{subfigure}
\hfill
\begin{subfigure}[t]{0.32\linewidth}
    \centering
    \includegraphics[width=\linewidth]{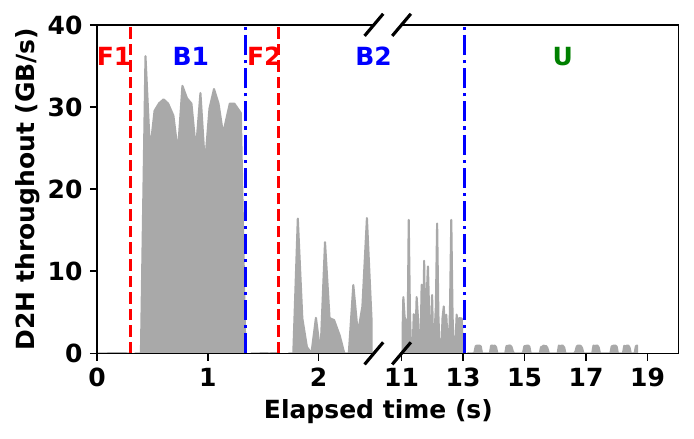}
    \caption{Device-to-host transfers.}
    \label{fig:d2h-throughput}
\end{subfigure}
\hfill
\begin{subfigure}[t]{0.32\linewidth}
    \centering
    \includegraphics[width=\linewidth]{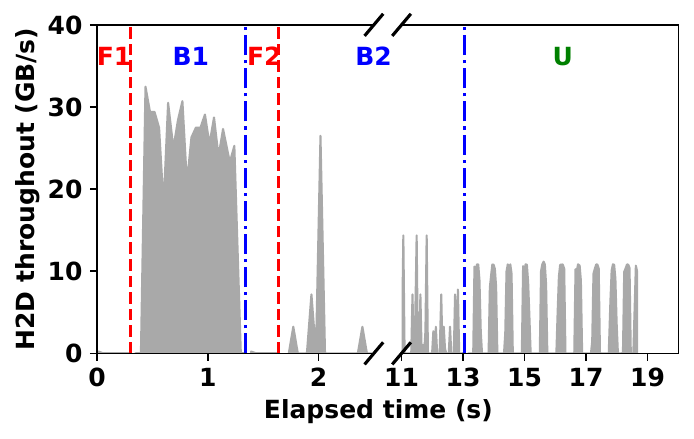}
    \caption{Host-to-device transfers.}
    \label{fig:h2d-throughput}
\end{subfigure}
\hfill
\caption{Memory and PCIe utilization statistics of single iteration when training 13B model.}
\label{fig:mem-pci-util}
\Description{
    A figure that shows the GPU memory utilization during different forward and backward phases. The second and third figure show the device-to-host and host-to-device transfer throughputs at different LLM training phases.}
\end{figure*}

\subsection{Experimental Setup}
To facilitate the study, we conduct our experiments on a single node of ALCF's JLSE testbed~\footnote{\url{https://www.jlse.anl.gov/}}
consisting of 4$\times$H100 GPUs with 80 GB HBM3 each (aggregated GPU memory of 320 GB), 2$\times$ Intel Xeon Platinum 8468 processors comprised of 48 CPUs each (total 96 cores, 192 threads), and 2$\times$ Gen4 NVMe of 1.5~TB each. The 512~GB DDR5 RAM is split across 2 NUMA domains. The peak unidirectional device-to-host~(D2H) and host-to-device~(H2D) throughput for pagable host memory are 16~GB/s and 9~GB/s respectively, whereas when the host memory is pinned, the D2H and H2D throughputs are both 55~GB/s. 
Our testbed runs PyTorch 2.2 with CUDA 11.8 and DeepSpeed 0.13.3 on top of the OpenSUSE Leap 15.4 operating system.

\subsection{Methodology}
Throughout our evaluations, we use the LLaMA2 13B model~\cite{touvronLlamaOpenFoundation2023}.
From the Bloom-175B~\cite{workshopBLOOM176BParameterOpenAccess2023} 
repository, we use the parameters of the Adam optimizer~\cite{ZeroInfinity-SC21},
the OSCAR-1GB~\cite{workshopBLOOM176BParameterOpenAccess2023}
dataset with a sequence length of 2048 tokens per microbatch, and a microbatch size of 1. DeepSpeed's ZeRO-3 engine~\cite{ZeroInfinity-SC21}
is used for training (which partitions the parameters, gradients, and optimizers across all data-parallel ranks to achieve the highest redundancy elimination). The optimizer state is completely offloaded to the CPU memory and uses the optimized Adam implementation from DeepSpeed- DeepSpeedCPUAdam~\cite{ZeroInfinity-SC21}.
The memory for storing the FP32 gradients is pinned on the host memory. To accelerate the computations during the backward pass, activation checkpointing is disabled, due to which activation recomputations are not needed. We evaluate all scenarios with gradient accumulation enabled, which is a widely adopted practice in the literature to minimize `frequent optimizer update' overheads. Therefore, we set the gradient accumulation degree to 2 (i.e., 2 microbatches will undergo their corresponding forward and backward passes before running a single update step) in order to understand the training behavior at the non-gradient boundary vs.\ the gradient boundary.

\subsection{Compute and Memory Breakdown Per Stage}
In our first set of experiments, we break down a single training iteration and systematically investigate the GPU compute, CPU compute, and memory utilization patterns for stages that contribute most significantly towards the overall iteration time. 

\subsubsection{Forward Pass:} 
As observed in Figure~\ref{fig:iter-time-breakdown}, the forward passes corresponding to the first and second mircobatches (labeled $F1$ and $F2$) constitute less than 4\% of the overall training time. This is unsurprising because the model is entirely stored on the GPU and no collective communication, e.g., parameter all-gather, is required because a single data-parallel rank is used, so all parameters are local GPU HBM residents. For DP=2 and DP=4, however, we observe a 23\% and 48\% higher forward pass times per microbatch due to parameter all-gather operations, even on high-speed NVLinks (140~GB/s) (see Table~\ref{tab:dp-scalability}). Therefore, larger models that are trained with higher degrees of data parallelism with ZeRO-3 incur performance penalties in forward pass due to parameter all-gather operations. 
Albeit, given the fact that backward passes are typically at least 2$\times$ computationally more intensive than the forward passes when all activations are stored, and incur the same parameter all-gather penalties, the fraction of forward pass would never constitute the major proportion of the training phase irrespective of the tensor and/or data-parallelism degrees.

\subsubsection{Backward Pass:}
When training with gradient accumulation enabled, multiple forward and backward pass pairs are executed before running a single update step, which enables larger ``effective" microbatch training. In this context, the forward-backward pair that are not immediately followed by an update step are termed as non-gradient accumulation boundaries ($B_N$ in Table~\ref{tab:dp-scalability}) ; while the backward passes immediately preceding an update step are termed as gradient boundary ($B_B$ in Table~\ref{tab:dp-scalability}).
The backward pass behaves differently for non-gradient accumulation boundaries as compared to the gradient accumulation boundary.
In particular, this distinction comes from the blocking D2H copy of accumulated gradients which happens at the last backward pass. These accumulated gradients are flushed to the CPU and are used by the offloaded CPU optimizer to run the update step. 

The backward phase and the all-reduce computations are performed in multiple buckets/chunks, which is determined by the \texttt{reduce\_bucket\_size} parameter of the DeepSpeed configuration (we used the default value of 0.5B). The \texttt{overlap\_comm} further allows overlapping gradient reduction of the previous bucket with the backward pass of the next bucket. For each all-reduce bucket, the following I/O and memory operations take place:

\textbf{$B^I$: Gradient accumulation buffer initialization:}
\label{sec:gradient-acc-initialization}
During the backward pass of the first microbatch, the host-resident gradient accumulation buffer is initialized with the GPU-resident gradient values of the first microbatch (D2H copy). To accelerate the accumulation operations (\texttt{tensor.add\_}), the gradient accumulation buffer corresponding to a single bucket is copied to the GPU (H2D). Then for each parameter that needs to be reduced, if its corresponding accumulation buffer exists on the GPU, the \texttt{grad\_acc\_buffer.add\_(grad\_val)} operation is invoked. Although these D2H and H2D transfers for creating the initial gradient accumulation buffer for a given bucket are non-blocking and overlap with the reduction process, the \textit{device to host data movements can be reduced by half} by simply doing a D2D copy of all-reduced gradients to initialize the GPU-resident accumulation gradient buffer and asynchronously copying the initial gradient values to the host-resident gradient accumulation buffer.

\begin{observationbox}
{\bf Observation 1:} During the backward pass of the first microbatch in an iteration, the redundant D2H and H2D copies can be eliminated when the initial GPU-resident gradient values are D2H copied to the host-resident gradient accumulation buffer.
\end{observationbox}

\textbf{$B^N$: Non-gradient accumulation boundary: }
\label{sec:non-gradient-boundary}
Starting from second microbatch onwards, for each bucket, if a gradient accumulation buffer corresponding to a parameter is not found on the GPU memory, the following operations are executed asynchronously on the default CUDA stream: (1) copy the gradient accumulation buffer from the host-resident gradient accumulation buffer to the GPU (H2D); (2) accumulate using \texttt{grad\_acc\_buffer.add\_(grad\_val)}; and (3) flush back the accumulated gradients to the host-resident gradient buffer (D2H).

\textbf{$B^B$: Gradient accumulation boundary:}
\label{sec:gradient-boundary}
During the last backward pass of the iteration, i.e., gradient accumulation boundary, in addition to running $B^N$, the gradients accumulated for all subgroups on the GPU buffer are D2H flushed to the host-resident gradient buffer, to be used by the CPU-resident optimizer during the update step, in blocking form. Therefore, the gradient reduction of the next bucket cannot overlap with the D2H transfer of accumulated gradients of the previous bucket. As observed in Figure~\ref{fig:iter-time-breakdown}, the time required for the last backward pass constitutes a major fraction of the overall iteration time.

\begin{observationbox}
{\bf Observation 2:} The backward pass can be accelerated by introducing asynchronous 
gradient flushes, at the expense of increasing the competition for PCIe 
links with other communications that are not using a different channel, e.g., when no NVLink is available.
\end{observationbox}

\begin{table}[t]
    \caption{Hybrid strong and weak scalability tests --- microbatch size remains the same per GPU, but model parameters get partitioned across DP ranks.}
    \label{tab:dp-scalability}
    \centering
    \begin{tabular}{|l||c|c|c|}
        \hline
        DP degree & 1 & 2 & 4 \\
        \hline \hline
        [$F$] Forward (s) & 0.31 & 0.37 & 0.39 \\
        
        [$B^N$] Backward non grad-boundary (s) & 1.03 & 0.82 & 0.7 \\
        
        [$B^B$] Backward grad-boundary (s) & 11.2 & 6.13 & 3.8 \\
        
        [$U$] Update (s) & 4.6 & 2.4 & 2.4 \\
        
        [$T$] TFLOPS & 18.9 & 32.6 & 23.4 \\
        \hline
    \end{tabular}
\end{table}

\begin{figure*}[t]
\centering
\begin{subfigure}[t]{0.32\linewidth}
    \centering
    \includegraphics[width=\linewidth]{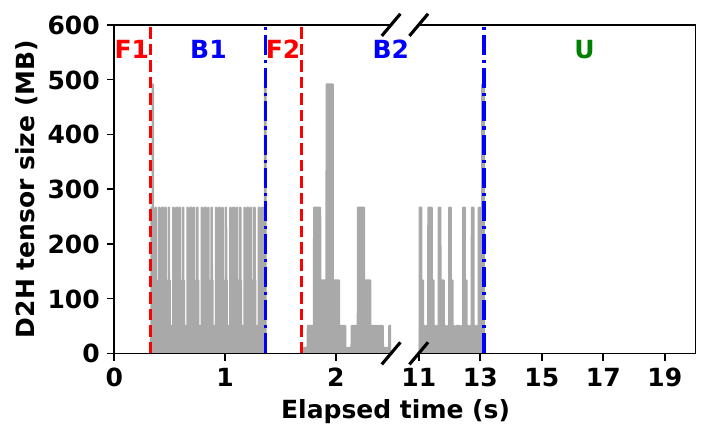}
    \caption{Tensor sizes during D2H transfers.}
    \label{fig:d2h-tensors}
\end{subfigure}
\hfill
\begin{subfigure}[t]{0.32\linewidth}
    \centering
    \includegraphics[width=\linewidth]{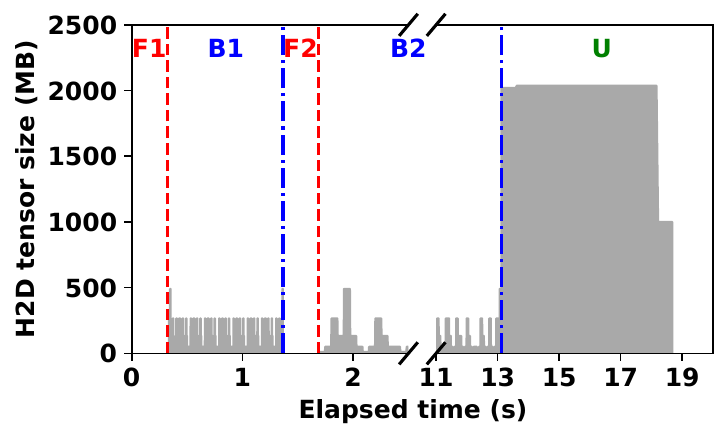}
    \caption{Tensors sizes during H2D transfers.}
    \label{fig:h2d-tensors}
\end{subfigure}
\hfill
\begin{subfigure}[t]{0.32\linewidth}
    \centering
    \includegraphics[width=\linewidth]{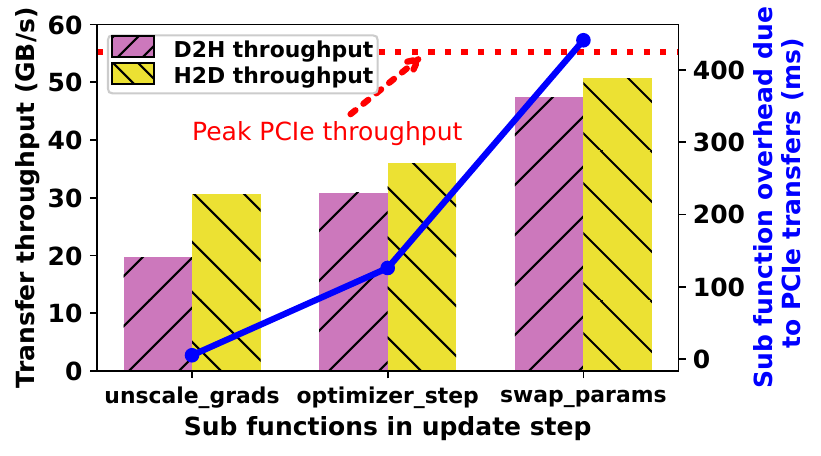}
    \caption{PCIe transfer throughput during different update subfunctions showing contention on host-memory.}
    \label{fig:breakdown-subfunctions}
\end{subfigure}
\hfill
\caption{Breakdown of tensor sizes transferred between device and host (Figure~\ref{fig:d2h-tensors} and Figure~\ref{fig:h2d-tensors}). PCIe transfer throughput during various subfunctions of the update phase (Figure~\ref{fig:breakdown-subfunctions}).}
\label{fig:tensor-sizes-update-breakdown}
\Description{
    A diagram that illustrates the sizes of tensors transferred from device-to-host and host-to-device during different training phases. Also, a diagram showing the the PCIe throughput decreases because of exhaustion of host-memory bandwidth when different subfunctions of the update phase are running.}
\end{figure*}

\subsubsection{Update Step:}
Given that the optimizer state is offloaded to the CPU, the update step runs completely on the CPU in a subgroup-by-subgroup fashion. The optimized DeepSpeedCPUAdam implementation of the Adam optimizer computes the update on the 96-core CPU at 4.5 billion parameters per second, whereas if the optimizer was completely executed on the GPU, it could run updates at 1 trillion parameters per second due to the embarrassingly parallel nature of the optimizer. However, as observed in the zoom of the update phase in Figure~\ref{fig:iter-time-breakdown}, the update computation of every subgroup is followed by H2D transfer of the updated FP32 parameters to the GPU-resident FP16 parameters (FP32 to FP16 conversion is done on the fly during transfer due to which only \texttt{sizeof(FP16)*num\_params} are transferred). As a consequence of such \textit{blocking the transfer of parameters}, the effective update throughput drops from 4.5 billion parameters per second to 2.8 billion parameters per second. When scaling the data-parallel degree, as shown in Table~\ref{tab:dp-scalability}, the update time [$U$] reduces to half for DP=2 due to parallel PCIe lanes available for transferring updated parameters concurrently to two different GPUs. However, for DP=4, the update time remains the same as for DP=2, demonstrating that the update computations on the oversubscribed 96-core CPU are a bottleneck rather than the PCIe transfer lanes.

\begin{observationbox}
{\bf Observation 3:} The effective throughput of the CPU-based optimizer updates drops by 60\% because of slow, blocking host-to-device transfer of updated parameters. 
\end{observationbox}

\subsection{GPU Memory and PCIe Link Utilization}
We also characterize the GPU memory utilization during various stages of training. Figure~\ref{fig:gpu-mem-util} shows the trend of GPU memory utilization reported by Nvidia Management Library (NVML)~\footnote{\url{https://developer.nvidia.com/nvidia-management-library-nvml}}
at 20~ms granularity throughout the training phases. However, we note that the PyTorch allocator caches the freed-up GPU memory for future reuse, and therefore the memory utilization shown in Figure~\ref{fig:gpu-mem-util} may not represent the exact quantum of utilized memory. Although the memory cached by the PyTorch allocator could be forcefully relinquished back to the GPU driver using \texttt{torch.cuda.empty\_cache()}, it is impractical to do so at fine-grained microsecond intervals because it leads to frequent memory allocation overheads. Furthermore, the major fraction of the GPU memory is allocated/deallocated at the forward, backward, and update stages because they work with tensors, which cannot be partially allocated or deallocated. Therefore, we only relinquish memory back to the system at the beginning of each forward, backward, and update stage, such that the impact due to relinquishing the cache is negligible on the iteration time. This approach allows us to broadly observe the growth of memory utilization at every stage. 

\subsubsection{Forward Pass:}
As observed from Table~\ref{tab:model-size}, the size for the 13B model consisting of FP16 params + FP16 gradients sums up to 46~GB. Because the gradients are not required during the forward pass, the memory utilization at the beginning of the forward pass for both microbatch-1 and microbatch-2 (represented as $F1$ and $F2$) in Figure~\ref{fig:gpu-mem-util}, sums up to $\sim$25~GB (some memory is utilized for storing CUDA and PyTorch contexts and GPU-based libraries). As the forward pass progresses, we observe $\sim$27~GB additional memory utilization to store the activations for a single microbatch, thereby showing high-memory pressure due to activations. 

We also observe, as shown in Figure~\ref{fig:d2h-tensors} and Figure~\ref{fig:h2d-tensors}, that no D2H or H2D transfers are made during the forward pass because all the activations are stored on the GPU memory itself for accelerating the backward pass. To alleviate the high-memory pressure of activations that are produced during the forward pass and consumed in the backward pass, we can enable activation checkpointing. This method can significantly reduce memory usage but will incur 33\% computation overhead due to the recomputation during the backward pass phase.

\subsubsection{Backward Pass:}
During the backward pass, the memory consumption stays nearly similar to that of the forward pass, as shown in Figure~\ref{fig:gpu-mem-util} (the broken x-axis between 2.5 and 11 has consistent memory utilization). This is because the gradients are accumulated bucket-by-bucket (defined by \texttt{reduce\_bucket\_size}) on the GPU and flushed back to the host memory. The sizes of tensors swapped in and out of the GPU memory during the backward pass are shown in Figure~\ref{fig:d2h-tensors} and Figure~\ref{fig:h2d-tensors}, respectively, which show high PCIe utilization at non-gradient accumulation boundaries, highlighted as region $B1$ in Figure~\ref{fig:d2h-throughput} and Figure~\ref{fig:h2d-throughput}.
At the gradient accumulation boundary (highlighted as region $B2$), we observe nearly similar sizes of tensors transferred from Figure~\ref{fig:d2h-tensors} and Figure~\ref{fig:h2d-tensors} (the broken x-axis has a similar trend as the latter part of $B2$). However, as observed in the region $B2$ of Figure~\ref{fig:d2h-throughput} and Figure~\ref{fig:h2d-throughput}, the D2H and H2D transfer throughput during the backward pass of the gradient accumulation phase is significantly lower than that observed during the non-gradient accumulation boundary ($B1$). The time used in region $B2$ accounts for 60.92\% of the overall time (i.e., $F1$+$B1$+$F2$+$B2$+$U$).
The transfers are slow (D2H throughput of 8~GB/s) because the Nvidia Nsight logs show that gradients are flushed to non-pinned memory on the host (even when the gradient buffers are pinned). This is because of mixed-precision training --- the gradients are copied in FP16 precision (in which the training is done) from GPU to the host memory in FP16 form (to unpinned buffer), and then converted from FP16 to FP32 (in the pinned buffer). 
While similar standalone tests for D2H tensor transfers showed near the peak PCIe utilization, the reason for slow D2H transfers during the last backpass is not completely understood. We suspect this may be because of synchronization issues between the asynchronous kernel executions on the GPU and the D2H transfers. 

\subsubsection{Update Step:}
Figure~\ref{fig:gpu-mem-util} shows that the GPU memory utilization during the update phase (labeled as region $U$), lowers down to just the FP16 model parameters, as is observed at the beginning of the iteration. This is obvious because the update process is done on the CPU, and no GPU memory is needed. Similarly, we do not observe traffic between the D2H during the update phase, as observed in Figure~\ref{fig:d2h-tensors} and Figure~\ref{fig:d2h-throughput}, because the gradients required for the update process have already been flushed during the last backward pass of the iteration. 
However, we observe significant H2D traffic in Figure~\ref{fig:h2d-throughput} due to flushes of large tensors (Figure~\ref{fig:h2d-tensors}), which correspond to the updated parameter values being copied from optimizer's master parameter copy (FP32) to the model's parameter copy. As shown in Figure~\ref{fig:iter-time-breakdown}, the zoom on the update phase shows that the computations on the CPU (denoted by the first row) do not overlap with the H2D transfers (second row) of the updated parameters. This highlights the opportunity to accelerate the update phase by concurrently running the H2D transfers of previous subgroups with the CPU-based updates of the next subgroups. However, in order to characterize this overlap opportunity, we need to analyze how memory-intensive the update phase is, and how the PCIe-based H2D transfers and CPU updates might compete for the limited bandwidth of the host memory.

\begin{observationbox}
{\bf Observation 4:} There are significant fluctuations of GPU memory utilization, i.e., high during
the forward and backward passes (due to accumulation of activations), 
and low during the update phase. Similarly, PCIe links are highly utilized 
during the backward passes but relatively less utilized during the update 
phase. There is a mix of small and large tensors being transferred.
This provides an opportunity for better host-device memory
management.
\end{observationbox}

\subsection{Host Memory Contention During Updates}
In our last set of experiments, we study how the subfunctions of the CPU offloaded optimizer update step impact the host-memory throughput.
Specifically, we analyze if the update operation is only CPU intensive (as can be seen in the zoom of the update step in Figure~\ref{fig:iter-time-breakdown}) or if it is memory intensive as well.
We identify three subfunctions that constitute the major fraction of the update step, namely \textit{unscale\_grads}, \textit{optimizer\_step}, and \textit{swap\_params}. To evaluate the host-memory contention, we first run each of these functions independently to obtain the time they take for execution. Then, we obtain the size of the tensor to be asynchronously transferred (over a dedicated CUDA stream) during each of these subfunctions by multiplying the peak PCIe throughput (55 GB/s) with the time required for each subfunction, such that the transfers can be perfectly overlapped with the subfunction execution on the CPU. The tensors used for testing the PCIe throughput are independently allocated on the host (pinned) and GPU memory such that it does not interfere with any of the existing buffers of the DeepSpeed engine. We repeat this experiment 5 times for each direction of transfer (H2D and D2H), and measure the transfer time using Nvidia CUDA profiler (NVTX)
events, and the slowdown incurred on the subfunction due to bandwidth contention on the host memory (represented by the blue solid line on the minor y-axis of Figure~\ref{fig:breakdown-subfunctions}). We observe from Figure~\ref{fig:breakdown-subfunctions} that the PCIe transfer throughput is drastically reduced ($\sim$55\%) when the gradients are being unscaled and clipped, which essentially performs a scalar multiplication operation with all elements of the subgroup gradients. Assuming fast scalar multiplication to the gradient tensor, the \textit{unscale\_grads} subfunction exhibits intensive read/write operations (of FP32 gradients) on the CPU memory, which results in slower PCIe transfers. 
Next, the CPUAdam optimizer step runs in an embarrassingly parallel fashion for all parameters of the subgroup, which involves read operation (for FP32 parameters, FP32 gradients, FP32 momentum, and FP32 variance), and write operation (for FP32 parameters, FP32 momentum, and FP32 variance). However, since \textit{optimizer\_step} is more CPU intensive, even with 7$\times$ FP32 read/write operations per parameter, we observe higher D2H and H2D throughput as compared to the \textit{unscale\_grads} subfunction. Lastly, the \textit{swap\_params} subfunction corresponds to copying the updated subgroup parameters from host to device (H2D). Here, although we observe significantly higher D2H and H2D throughputs for our test tensors (which are pinned on the host memory), the penalty for these faster transfers is paid by the slowdown of the subfunction itself (denoted by the blue solid line associated with the minor y-axis). Therefore, all subfunctions demonstrate some degree of host memory intensiveness due to which asynchronous D2H and/or H2D transfers cannot be run at peak PCIe throughput. 

\begin{observationbox}
{\bf Observation 5:} The CPU offloaded optimizer consists of memory-intensive operations, resulting in 
lower host memory bandwidth utilization
for device-host transfers across the PCIe link due to contention for the host memory bandwidth.
\end{observationbox}

\section{Conclusions}
\label{sec:conclusions}

In this work, we analyze how forward passes, backward passes and
update steps interleave to exhibit different behaviors during a training 
iteration of LLMs and transformers using a state-of-the-art runtime such as 
DeepSpeed.
Given the explosion of model sizes and optimizer states,
we have focused specifically on ZeroInfinity,
an offloading solution that reduces the GPU memory utilization by placing
partitions of the optimizer state and model parameters on the host memory, 
where the update step is computed directly on the CPU. 
Our study has revealed
several important bottlenecks of existing implementations: redundant 
and/or blocking device-to-host and host-to-device data transfers that 
slow down the last backward pass and the update step, slow upload of
updated model parameters back to the GPUs in half precision, fluctuations
of GPU memory utilization and PCIe links that lead to underutilization
of GPUs especially during the update step, contention for PCIe links
due to 3D parallelism, and contention for the I/O bandwidth of the 
host memory due to concurrent data transfers and CPU computations. 
Encouraged by these observations, in future work we envision designing
and developing flexible hybrid GPU-CPU offloading solutions that adapt
to the fluctuations of GPU memory and PCIe link utilization to
reduce the aggregated GPU memory footprint as much as possible, 
while minimizing the scalability and performance penalty of doing so.
Specifically, we predict two opportunities. First, during the last
backward pass preceding the update step, as the activations are
gradually deallocated after computing the corresponding gradients,
the spare GPU memory can be used to buffer the gradients and asynchronously
flush them to the host memory, thereby significantly accelerating the 
backward pass. These asynchronous flushes can take advantage of 
different tensor sizes to coalesce and prioritize data transfer operations 
to optimize the PCIe link utilization without slowing
down other communications during the backward pass that also share 
the same PCIe link. Second, during the update step the GPUs
are underutilized, both in terms of memory and compute capability.
However, unlike state-of-the-art approaches that use a fixed ratio
to partition the optimizer state and model parameters among the GPUs and
the host memory, we envision a flexible solution that adapts to the
fluctuations of GPU memory utilization to temporarily
offload during the update step as many computations as possible
on the GPUs, and then move the updated parameters back to the 
host memory to make room for the activations of the next forward
pass.

\section*{Acknowledgements}
This work is supported in part by the U.S. Department of Energy (DOE), Office of Science, Office of Advanced Scientific Computing Research under contract DEAC02-06CH11357/0F-60169 and the National Science Foundation (NSF) under award no.\ 2106634/2106635. Results presented in this paper are obtained using Argonne's ALCF HPC systems, and NSF Cloudlab and Chameleon testbed.

\balance

\bibliographystyle{IEEEtran}
\bibliography{references}

\end{document}